\documentclass{aa}
\usepackage{psfig} 
\begin{document}

\title{On the Nature of X-ray Variability in Ark~564}
  
\author{M. Gliozzi\inst{1}\inst{,2}
\and  W. Brinkmann\inst{3} 
\and C. R\"ath\inst{3}
\and I.E. Papadakis\inst{4}\inst{,5}
\and H. Negoro\inst{6}
\and H. Scheingraber\inst{3}} 
\offprints{mario@physics.gmu.edu} 
\institute{
George Mason University,
Department of Physics and Astronomy, MS 3F3, 4400 University Dr., Fairfax, 
VA 22030-4444
\and Max-Planck-Institut f\"ur extraterrestrische Physik,
         Postfach 1312, D-85741 Garching, Germany
\and Centre for Interdisciplinary Plasma Science,  Max-Planck-Institut f\"ur 
extraterrestrische Physik, Postfach 1312, D-85741 Garching, Germany 
\and Foundation for Research and Technology--Hellas, 711 10 Heraklion, Crete,
Greece      
\and Physics Department, University of Crete,  
710 03 Heraklion, Crete, Greece
\and Cosmic Radiation Laboratory, RIKEN, 2-1 Hirosawa, Wako--shi,
Saitama 351-0198, Japan
}

\date{Received: ; accepted: }

\abstract{We use data from a recent long ASCA observation of
the Narrow Line Seyfert 1 galaxy \object{Ark~564} to
investigate in detail its timing properties.
We show that a thorough analysis of the time series, employing 
techniques not generally applied to AGN light curves,
can provide useful information to characterize the engines of these
powerful sources.
We searched for signs of non--stationarity in the data, but did not 
find strong evidences for it. 
We find that the process causing the variability is very likely nonlinear,
suggesting that variability models based on many active regions, as the
shot noise model, may not be applicable to \object{Ark~564}.
The complex light curve can be viewed, for a limited range of time scales
(as indicated by the breaks in the structure and power density spectrum), 
as a fractal object 
with non--trivial fractal dimension and statistical self--similarity.
Finally, using a nonlinear statistic based on the scaling index as a tool to 
discriminate time series, we demonstrate that the high and low count rate
states, which are indistinguishable on the basis of their
autocorrelation, structure and probability density functions,
are intrinsically different, with the high state characterized by  
higher complexity.
}
\titlerunning{Nature of X-ray variability}
\authorrunning{M.~Gliozzi et al.}
\maketitle
\section{Introduction}
Active Galactic Nuclei (AGN) are variable in every observable wave band.
The X-ray flux exhibits variability on time scales shorter than any other
energy band, indicating that the emission occurs in the innermost regions
of the central engine. Therefore, a study of the X-ray variability 
provides an additional 
powerful tool to probe the extreme physical processes operating in the inner 
parts of the accretion flow close to the accreting black hole. Although X-ray 
variability has been observed in AGN for more than two decades, its origin 
and nature is still poorly understood.
Until recently, there were no sufficiently long observations  
of AGN with good signal to noise to allow a thorough analysis.
Secondly, the temporal analysis is frequently limited to 
calculations in the Fourier domain of the power density spectrum (or, 
equivalently, in the time domain of the structure function) which, although 
useful for detecting possible periodicities or typical time scales in a signal, 
does not exploit all the potential information contained in a time series. The
reason is that this technique uses only the first two moments (namely the mean
and the variance) of the probability distribution function associated with the 
physical processes underlying the signal. 
However, only a Gaussian distribution can be
completely described by the first two moments, and there are several indications
that probability density functions associated with X-ray light curves both in
AGN and in Galactic black hole systems, are not Gaussian 
(e.g. Leighly \cite{leig1}, Greenhough \cite{green}).
Finally, the light curves,
i.e. the starting point for any kind of temporal analysis, have been often
considered as a by-product of the spectral analysis, which still catalyzes 
most of the attention and efforts, in particular now that high resolution 
X-ray spectroscopy of AGN is possible thank to {\it Chandra} and 
{\it XMM-Newton}. Spectral analysis proved to be very 
useful in providing constraints on the physical parameters of the accretion
 flow around black holes but it must be pointed out that, 
due to low signal-to-noise data, spectral models are often applied to 
time-averaged spectra, despite the fact that sources show rapid 
variability. Thus the additional information provided by timing observations
can be crucial to break the degeneracy among spectral models.

One of the most critical open questions, related to the X-ray variability in 
AGN, concerns the nature of the variability: is it linear or nonlinear? 
In a mathematical sense linearity means that the value 
 of the time series at a given time 
can be written as a linear combination of the values at previous times 
plus some random variable.
A positive detection of nonlinearity  
would have an immediate and very important consequence for the modeling of
the region producing X-rays: all the variability models based on many
independent active regions, as the shot noise model (Terrel 1972) or 
magnetic flares (e.g. Galeev et al. 1979) would be ruled out in favor of
inherently nonlinear models as the self-organized criticality disk model
(Mineshige et al. 1994) or the emission of X--ray radiation from a putative jet.

\begin{figure}
\psfig{figure=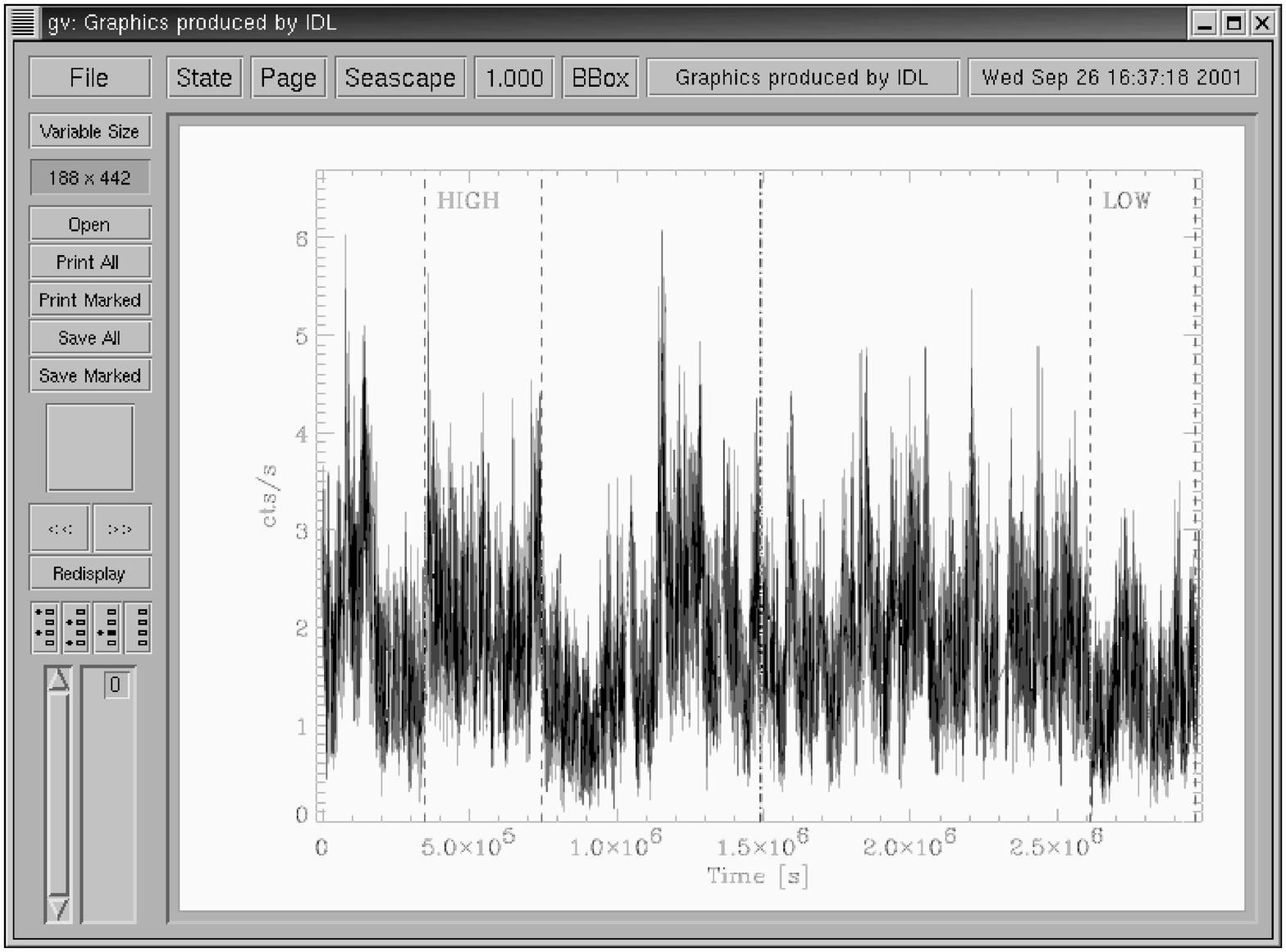,height=6.5cm,width=8.7cm,%
bbllx=124pt,bblly=153pt,bburx=692pt,bbury=581pt,angle=0,clip=}
\caption{ASCA SIS0 light curve of \object{Ark564}. The dot--dashed thick
line divides the light curves in two halves used to investigate 
stationarity. The dashed lines define two intervals used to characterize the
timing behavior during the low and high count rate states.
\label{figure:lc-ark}}
\end{figure}
In an attempt to answer this question and, more generally, to 
investigate the nature of the X-ray variability, we focus on 
a prominent object of a particular 
class of AGN, the Narrow-Line Seyfert 1 galaxies (NLS1), which often display 
rapid, large amplitude X-ray variability as well as extreme long-term changes
(Forster \& Halpern \cite{forst}, Boller \cite{boll2}, Brandt \cite{brand2}), 
and therefore represent the ideal objects for an X-ray temporal analysis. In 
particular we analyze a recent long ASCA observation of \object{Ark~564}, 
the brightest NLS1 in the 2-10 keV band, using non-standard (at least 
for X-ray astronomy) analysis techniques.
 
The outline of this paper is the following. In Sect. 2 we describe the
data used for the timing analysis. In Sect. 3 we investigate the important
issue of the stationarity of a time series, by dividing the light curve into two
equal parts and by computing and comparing the mean, the variance, the 
auto-correlation and structure functions and the power spectra of the two 
halves. Sect. 4 deals with the search for nonlinearity. To this end, we choose
two parts of the light curve with reasonable length ($\sim$ 4 days), when
the source was at a high and low count rate state, respectively. We use a 
new technique based on the constrained randomization of a time series and 
the nonlinear prediction error as an indicator of nonlinearity,
in order to test the hypothesis that the light  curve is a realization of a
linear process.
In Sect. 5 we first investigate the variability behavior of the source,
utilizing both standard (excess variance, probability density function) and
non--standard (fractals) techniques. Then, in order to investigate further
whether the behavior of the source is identical in the two states (high and
low) we used techniques like the phase space reconstruction and the
scaling index method. Finally, in Sect. 6 we draw the main conclusions.

\begin{figure}
\psfig{figure=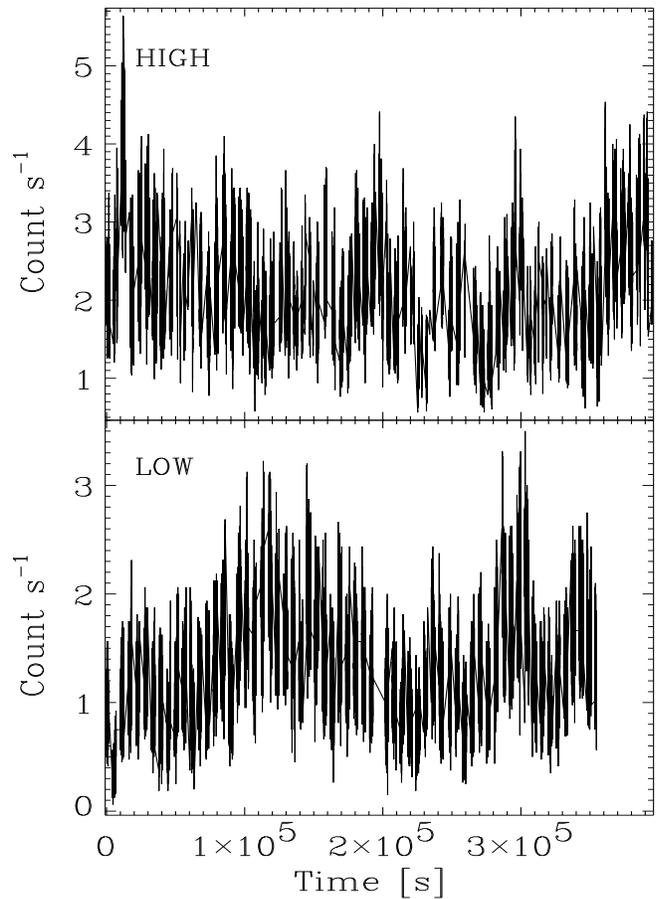,height=12cm,width=8.7cm,%
bbllx=44pt,bblly=29pt,bburx=370pt,bbury=567pt,angle=0,clip=}
\caption{ASCA SIS0 light curves of \object{Ark564} during the high 
(top panel) and the low (bottom panel) count rate state. Time bins are
16s. Both light curves contain 10000 data points each.
\label{figure:lc2a}}
\end{figure}

\section{Data}
\object{Ark~564} was observed by ASCA from 2000 June 01 11:51:27 to 2000 July 
05 23:57:54. We extracted the X-ray data from the public archive at ISAS 
and applied standard criteria for the data analysis in order to create 
light curves. Source counts were accumulated from a $\sim
8\arcmin \times 8\arcmin$ rectangle in detector coordinates to maximize
the X-ray signal for the SIS0. The background was estimated from the source
free region on the same chip and its level
was less than 3\% of the source count rate.
During the analysis it became apparent that all instruments suffered from
unexpected light leaks. This is probably due to an expansion of the atmosphere
caused by recent solar activity and to the low satellite orbit during the
observation. We investigated this light leak effect using energy spectra, low
energy light curves,  and {\tt DFE} curves for each of the individual data sets
(each {\tt FRF} file). We found that in the data screening the {\tt BR\_EARTH}
criterion for the SIS data had to be changed from $20^{\rm o}$ to $35^{\rm o}$
for the 0.55--10~keV data.
The background-subtracted light curve is plotted in Fig.~\ref{figure:lc-ark}.
The reason for
the selection of two sub-intervals with high and low mean count rate
is explained in Section 3. Fig.~\ref{figure:lc2a} shows a blow-up of the
light curves during the high (top panel) and low count rate state (bottom panel).
In the following, we restrict our analysis to the energy range 0.7--10 keV,
completely free of the light leak problem, and use data only
from the SIS0 detector, which is the most sensitive (and therefore has the
highest count rate). 

\section{Stationarity Check}
A time series, as any other scientific measurement, has to provide enough 
information to allow the determination of the quantity
of interest unambiguously and it is
useful only if reproducible, at least in principle. In other words, all  
statistical properties derived from the time series analysis should be
invariant under a shift of the time origin. This is the definition
of a completely stationary process.

However for most of the physical processes, in particular in astrophysics,
this severe requirement is not applicable and two weaker conditions (which
nevertheless describe roughly the same physical behavior) are applied. 
As first requirement, the time series should cover a stretch of time which
is much longer than the characteristic time scale of the system. A second
important requirement is that the properties of the system generating the 
signal must not change during the observation period. This can be checked in
principle by measuring statistical properties, as the mean and 
the variance 
for the first and the second half of the data available and verifying that
their variations are within their statistical fluctuations.

However, in case of AGN X-ray light curves, these requirements are 
usually not fulfilled for the following two reasons. First, X-ray observations
of AGN have typically a duration ranging
between a few hours to a few days (in the best cases). This time interval is 
usually sufficient to accumulate energy spectra with good signal-to-noise ratio, 
but not to reveal the characteristic time scale of an AGN. Second,
the AGN power density spectrum is characterized  by a ``red noise" power law
component $P(f)\propto f^{-\alpha}$ (with $\alpha\sim
1.5$; e.g. Lawrence \& Papadakis 1993), indicating the presence of temporal
correlations. Therefore, even if the parameters of the
power density spectrum do not change, other parameters as the mean can change. 
However, as demonstrated by Press (1978) and, more recently
for NLS1 galaxies, by Leighly (\cite{leig1}),
the differences in excess variance between the first and the second half of 
the light curve can be entirely ascribed to the weak nonstationary inherent 
in the red noise. 
\begin{table} 
\caption{Statistical comparison between first and second half of the light curve}
\begin{center}
\begin{tabular}{lll}
\hline
\hline
\noalign{\smallskip}
Property & First half& Second half\\
\noalign{\smallskip}       
\hline
\noalign{\smallskip}
\noalign{\smallskip}
Mean & $1.893\pm 0.004$ & $1.727\pm 0.003$\\
\noalign{\smallskip}
\hline
\noalign{\smallskip}
Variance & $0.600\pm 0.004$ & $0.440\pm 0.003$ \\
\noalign{\smallskip}
\hline
\noalign{\smallskip}
Skewness & $0.663\pm0.013$ & $0.625\pm0.013$\\
\noalign{\smallskip}
\hline
\noalign{\smallskip}
Kurtosis & $ 0.732\pm0.026$ & $ 0.696\pm0.025$\\
\noalign{\smallskip} 
\hline
\end{tabular}
\end{center}
\end{table}
\begin{figure}
\psfig{figure=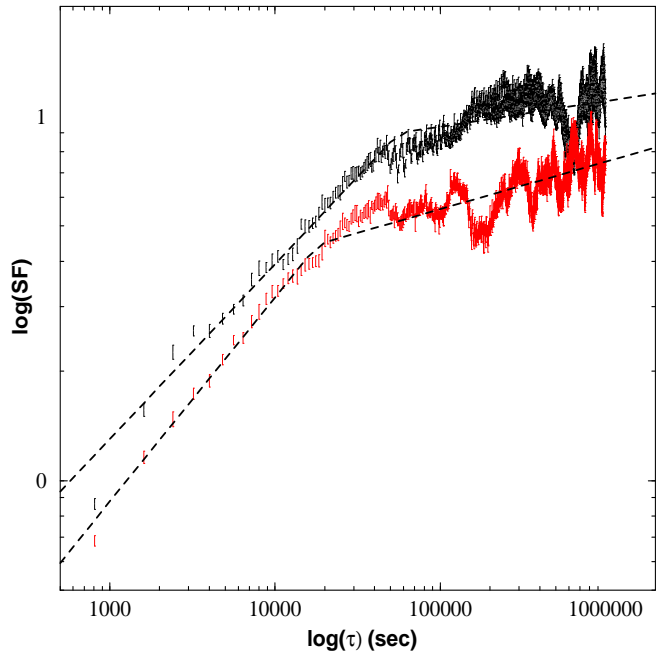,height=8.7cm,width=8.7cm,%
bbllx=65pt,bblly=278pt,bburx=510pt,bbury=703pt,angle=0,clip=}
\caption{Structure function of \object{Ark~564}. The black error bars 
(upper curve)
correspond to the first half of the light curve, the gray ones (lower curve)
to the second half. The dotted lines show the best fitting ``broken power law"
model to the data.
\label{figure:SF-ab}}
\end{figure}

\begin{figure}
\psfig{figure=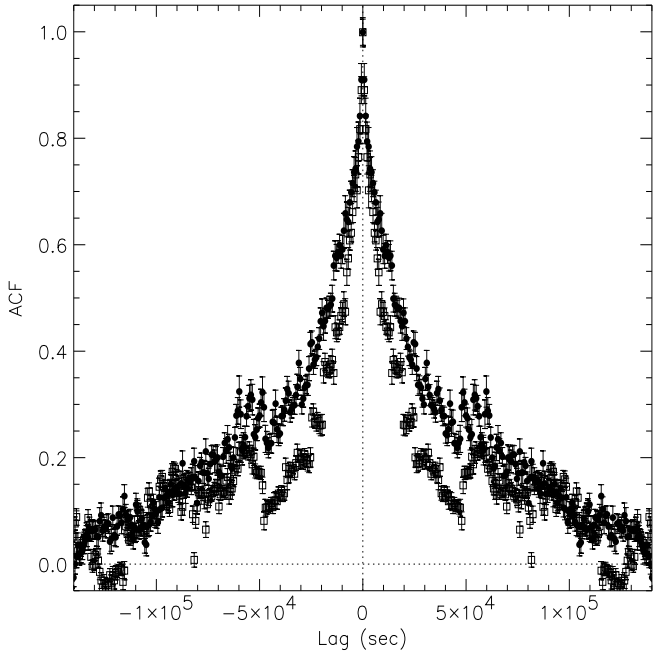,height=8.7cm,width=8.7cm,%
bbllx=60pt,bblly=122pt,bburx=574pt,bbury=600pt,angle=0,clip=}
\caption{Autocorrelation function of \object{Ark~564}. 
The filled circles correspond
to the first half of the light curve, the open squares to the second half. 
\label{figure:ACF-ab}}
\end{figure}

The long light curve of \object{Ark~564} allows an accurate investigation of
the stationarity of the source. For this reason, we divided the light curve
into two equal parts, and computed not only the usual parameters, as the mean
and the variance, but also the skewness and kurtosis (listed in Table 1) and
estimated the structure and auto--correlation functions of the two halves.

The errors quoted in Table 1 were computed assuming that
the data distribution is normal and the data are a collection 
of independent measurements.  
However, in the time series considered (and in general in all the AGN light 
curves)
the measurements are not independent, but correlated. In this case, 
the actual errors, which might be much
larger than the errors computed as explained above, cannot be easily quantified,
since they depend on unknown quantities. For example, the actual error of the 
observed mean and variance depends on the true variance, on the true 
autocorrelation function of the underlying process
and on the value of the power density spectrum at zero frequency 
(see Priestley 1989).
Therefore, since the difference of the mean and variance between the two
halves could be the result of the fact that red-noise processes are weakly
non-stationary, we consider this difference as an indication, not as a
proof, for stationarity.

To further investigate this issue we performed an analysis based on
the structure function (e.g. Simonetti \cite{simo}, Hughes \cite{hug}),
which has the ability to discern the range of time scales that 
contribute to the variations in the data set.
It is well known that structure functions
are affected by the strong correlation between the points in the light
curve. As with other statistics in the time domain, like auto and cross
correlation functions, the points in the structure function are not
independent, and the degree of their dependence is not easy to be
determined in advance. However, the structure functions are not affected
by the presence of missing parts in the light curves, therefore, their use
can provide us with useful information in the case of unevenly sampled
data, like in our case. For that reason, we computed the structure
functions for the first and second half of the light curve. The results
are plotted in Fig~\ref{figure:SF-ab} (black and grey curve, respectively).
The most important difference in 
Fig~\ref{figure:SF-ab} is the 
time at which the structure functions reach a plateau state. By fitting a
broken power law model to both structure functions, the best fit
break time scale is $5.6^{+2.7}_{-1.8}\times 10^4 {~\rm s}$
for the first part of the light curve and 
$1.8^{+1.2}_{-0.8}\times 10^4 {~\rm s}$
during the second interval (the errors correspond to a 68\% confidence level).
This intrinsic difference in the temporal properties of the two halves of the 
light curve seems also to be reflected in the difference of 
the auto--correlation functions, estimated by using the
Discrete Correlation Function (DCF) of Edelson \& Krolik (1988).
The rate at which the auto--correlation function decays to zero may
be interpreted as a measure of the ``memory" of the process and thus 
Fig~\ref{figure:ACF-ab} 
indicates that the ``memory" of the second part of the light curve is shorter
than that of the first part. 

Like the mean and variance, the structure function and autocorrelation
function of the first and second half of the light curve show
differences. However,
before reaching the conclusion that the light curve of \object{Ark~564} is 
non--stationary, we should caution that it is very difficult to determine
appropriate errors for the structure and autocorrelation functions
and thus to assess the statistical significance of the apparent difference.
(the reported errors basically give information about how many 
points contribute to each time scale.)
We therefore tried to
investigate further the stationarity using the power density
spectrum (e.g. Papadakis \& Lawrence \cite{pap2}) which, being the counterpart
of the structure function in the Fourier domain, should provide the same
information. 
In fact, detailed analysis of the power spectrum of the whole ASCA light
curve of Ark 564 (Papadakis et al. 2002), together with the results of
Pounds et al. (2001), shows that the power spectrum actually starts to
flatten below $\sim 10^{-5}$ Hz, in agreement with the structure functions
plotted in Fig~\ref{figure:SF-ab}. However, things are a bit more complicated 
when we divide the total light curve in two parts and compute their power 
spectra.  
The power spectrum is a much ``noisier" function, when compared to the
structure function. As a result, significant binning of the ``raw" power
spectrum is needed in order to investigate its shape. Without going into
details of how the power spectrum was computed (see Papadakis et al. 2002),
we simply mention that for a time series of 15 days length (corresponding to
a minimum frequency of $7.7\times 10^{-7}$ Hz), after the necessary
binning of the spectrum, the lowest accessible frequency is higher than
$10^{-5}$ Hz. Therefore, all information about the possible presence
of a break above $10^{5}$s is inaccessible and we cannot investigate if the
location is different between the two parts of the light curve. Nevertheless,
the overall power spectra of the two parts are consistent within the
errors, a result that argues in favor of the the light curve being
stationary.

Concluding, we can say that we find no strong evidence for
non-stationarity. The mean, variance, structure function and
autocorrelation functions of the two parts of the light curve do show
differences, but these could be the result of the red-noise character of
the observed variations. In fact, the power spectra of these parts show no
statistically significant differences.
However, since non-stationarity can affect the identification of genuine
nonlinearity present in the data, and because of the presence of
indications for non-stationarity (although not statistically
significant) we decided not to utilize the whole data set for an  
investigation of the question of linearity of the light curve. Instead,
we chose two parts that are ``locally" stationary (meaning that after dividing
them in two halves, no significant difference was found between the mean, 
variance, autocorrelation and structure functions of the first and second part),
have reasonable length ($\sim 4$ days), and have different mean count rates.
The selected ``high" and ``low" intervals, plotted in Fig~\ref{figure:lc2a}, 
contain 10000 data points each (using a bin size of 16 s) with mean 
count rates of $2.14\pm 0.01{\rm~cts~s^{-1}}$ and 
$1.32\pm 0.01{\rm~cts~s^{-1}}$, respectively. In this way, apart from checking 
whether the light curve is nonlinear or not, we can also test if the source 
behaves differently in different flux states. 

\section{Search for nonlinearity}
The search for nonlinearity in a time series is not only important but
unavoidable in order to progress in the understanding of the origin
of X-ray variability in AGN. In fact, various kinds of models (Gaussian, 
non-Gaussian, or nonlinear) have been proposed to explain the AGN variability.
All of them are able to explain the red--noise power spectrum and other temporal
characteristics such as the auto- and cross-correlation functions. However, it 
is important to stress that neither the power density spectrum nor the 
auto-/cross-correlation or structure functions are able to distinguish whether 
a process is linear or nonlinear (e.g. Vio et al. 1992).
\begin{figure}
\psfig{figure=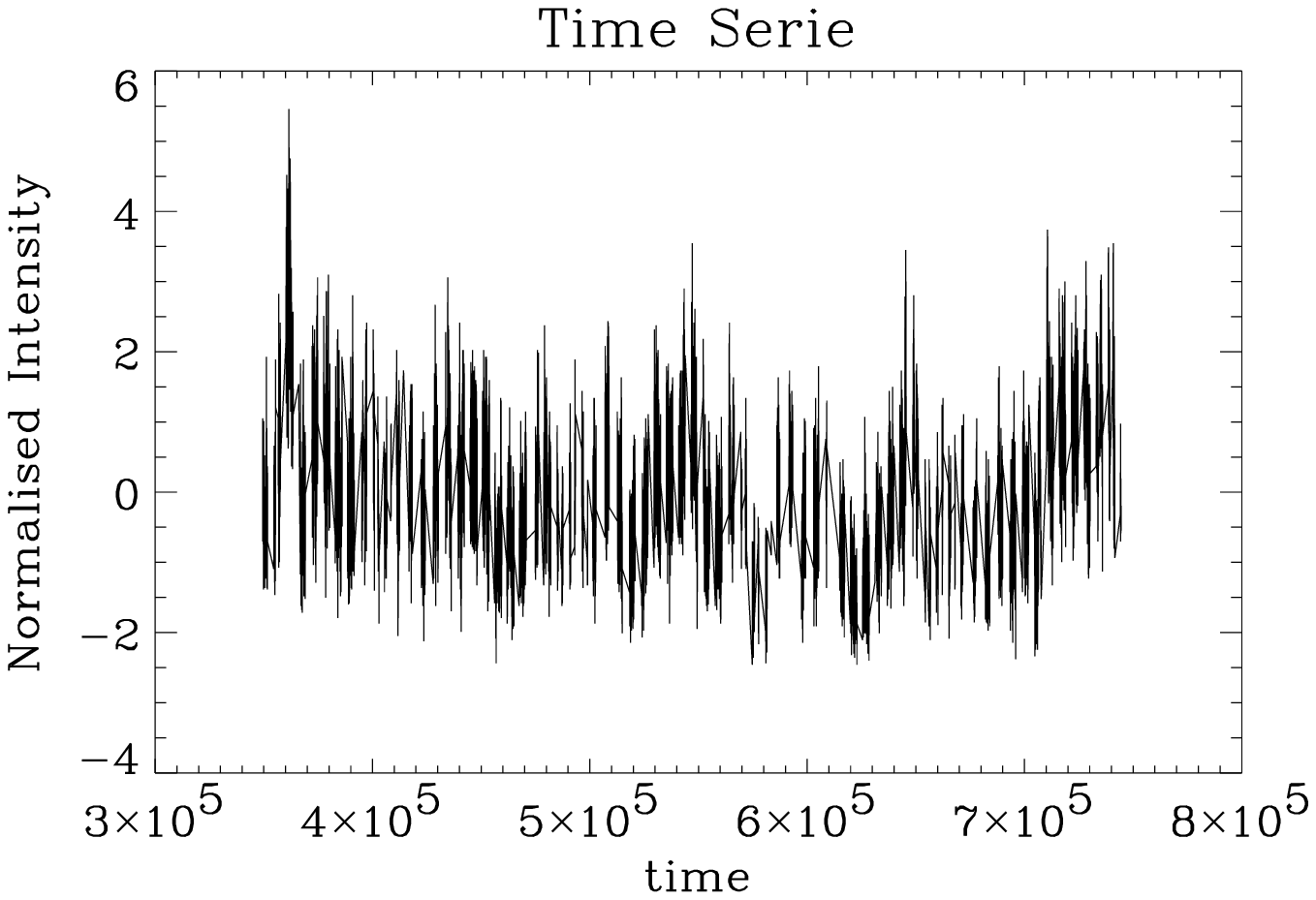,height=5cm,width=8.7cm,%
bbllx=30pt,bblly=10pt,bburx=426pt,bbury=261pt,angle=0,clip=}
\psfig{figure=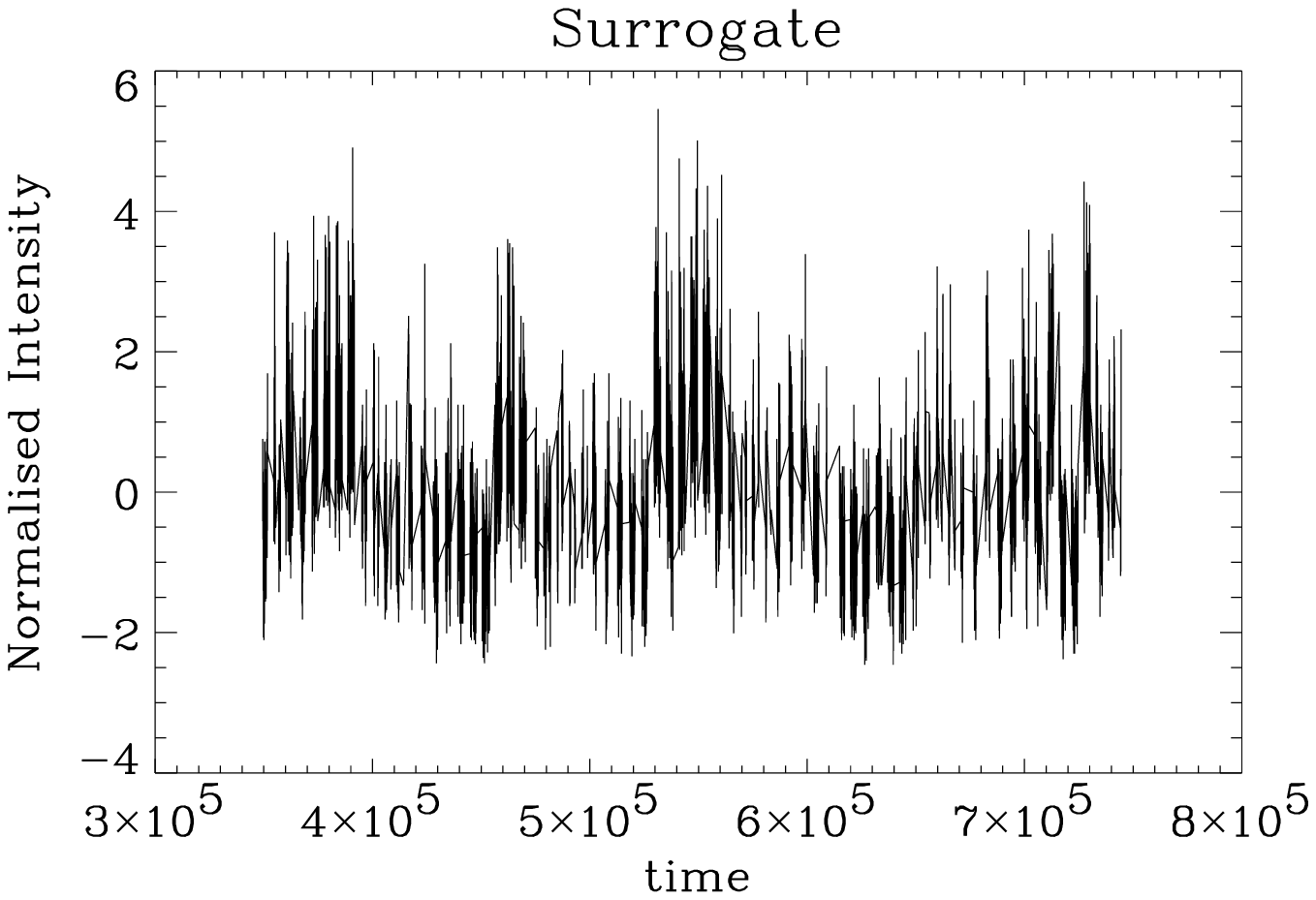,height=5cm,width=8.7cm,%
bbllx=30pt,bblly=10pt,bburx=425pt,bbury=261pt,angle=0,clip=}
\caption{Normalized real time series (top panel; mean = 0, $\sigma = 1$)
 for the high state of 
\object{Ark~564} and corresponding surrogate time series (bottom), which
preserves the linear properties of the original data as mean, variance and
auto--correlation function.
\label{figure:surrog}}
\end{figure}

A relatively simple but sensitive method to test the presence of nonlinearity
in a signal is the so called method of surrogate data (e.g. Theiler et al.
1992), which is based on the following ideas. First, one has to conceive
a null hypothesis (for example, that the surrogates have the same linear 
properties as the real data) that we want to test against.
Second, a level of significance for the test must be specified (a test valid
at 95\% significance level means that with a chance of $\alpha=0.05$ the
null hypothesis is erroneously rejected). Then one 
creates a number of surrogate data according to the chosen null hypothesis
( $N_{\rm surr}=1/\alpha-1$ are created for a one--sided test, where the
null hypothesis is rejected if the data deviate from the surrogates in a
specified direction, whereas $N_{\rm surr}=2/\alpha-1$ are created in case of
a two-sided test, where the data can deviate from the surrogates on either side).
Finally one has to compare the real data with the surrogate time series using
nonlinear statistics. If the real data differ
significantly from the surrogates, nonlinearity can be inferred.

Standard surrogate methods (Theiler et al. 1992), make use of the Fourier 
transformation to conserve
the auto--correlations of the original data: the original time series is 
rescaled to a Gaussian distribution first; then, the Fourier phases are 
randomized and the Fourier transformation is inverted. Finally, a rescaling to
the original distribution is performed. However, such a method cannot be
applied to unevenly sampled data (as the ASCA light curve of \object{Ark~564}),
because it utilizes Fourier transformations
and their inverses. 
Therefore, a more general approach based on the constrained randomization
of a time series (Schreiber 1998) was used. In this new method,
the desired properties (e.g. auto--correlations) of the surrogates are 
imposed by constraints: the surrogates start as a randomized
distribution of the original data, and then are annealed until they match 
the ACF of the original data, at which point they are accepted as surrogates.
The constraints are implemented as a cost function
$E$ (e.g. Schmitz \& Schreiber 1999), which is constructed to have a global
minimum when the constraint is fulfilled (in simple words, if the constraint
chosen is that surrogate data have the same auto--correlation as the real data,
$E$ will be a function of the difference between the auto--correlations
of real and surrogate data). Combinatorial minimization by complete enumeration
is not feasible in case of long time series, since the computational effort
grows exponentially with the length of the time series. Instead, we use
the so called simulated annealing (e.g. Metropolis et al. 1953, Kirkpatrick
et al. 1983), a powerful method for minimization in presence of many false
(local) minima, that is expected to find an approximate solution in polynomial 
time.
\begin{figure}
\psfig{figure=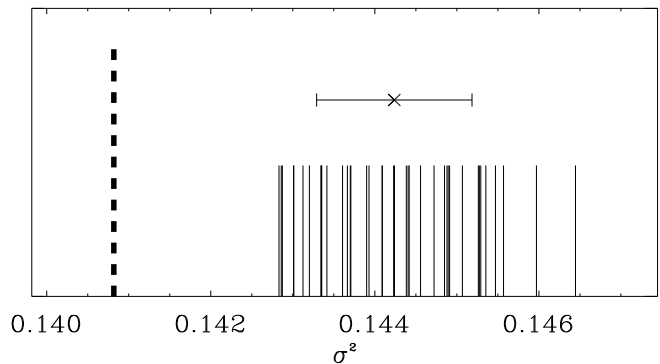,height=5cm,width=8.7cm,%
bbllx=66pt,bblly=6pt,bburx=405pt,bbury=205pt,angle=0,clip=}
\caption{Nonlinear prediction errors calculated from the surrogate data
(shorter lines) and from the real data (longer thick dashed line) during the high
state. The
horizontal error bar indicates the standard deviation around the mean of the
statistic computed from the surrogates.
\label{figure:result2}}
\end{figure}

The main idea behind the method of simulated annealing is to interpret the
cost function $E$ as the energy in a thermodynamic system. Minimizing $E$ then is
equivalent to finding the ground state of the system, and typically, for a 
glassy solid, this state is reached by first heating and subsequently cooling 
it (a procedure called ``annealing", hence the name of the method). In order to
anneal the system to the ``ground state", that is, to the minimum cost function,
we first ``melt" the system at a high temperature $T$, and then decrease
$T$ slowly. In practice, we started with a random permutation of the original 
time series and defined the starting cost function in the following way:
\begin{equation} 
E=\left\{\sum_{i}\left[ACF_{\rm real}(i)-ACF_{\rm surr}(i)\right]^2\right\}^{1/2},
\end{equation}
where $ACF_{\rm real}$ is the binned auto-correlation function of the real data
and $ACF_{\rm surr}$ of surrogates. The surrogate is
successively modified by exchanging randomly chosen pairs of elements and thus
also the value of the cost function changes. The modification will be accepted
with a probability $p=1$ if the energy decreases ($\Delta E<0$) and with
$p=\exp(-\Delta E/T)$, where $T$ is a free parameter, if  $\Delta E \geq 0$.
Such updating scheme, proposed by Metropolis and collaborators (1953), allows the
system to be close to the ``thermodynamic equilibrium" at each stage.
To summarize: the simulated annealing is performed by exchanging 
pairs of points (in the time domain) in each iteration step.
Therefore, the surrogates have exactly the same distribution
as the real data, i.e. they are ``amplitude adjusted". 
By keeping their autocorrelation function close to that of the original 
data all linear
properties of the initial time series are preserved.

We repeated the whole procedure several times and created 39
independent surrogate data sets, 
necessary to perform a two--sided test for nonlinearity valid at 95\%
significance level or a one-sided test at 97.5\% significance level
(see below).  Fig~\ref{figure:surrog} shows the comparison
between the real normalized time series (normalized to a mean$ = 0$ and  
$\sigma = 1$) during the high state of 
\object{Ark~564} and a corresponding surrogate time series (bottom), which
preserves the linear properties of the original data like mean, variance and
auto--correlation function. 

After having produced randomized versions of unevenly sampled time series with
given linear correlations, we use such surrogates to test for 
nonlinearity in the original data. In principle, any nonlinear statistic might
be used, however, many statistics useful for evenly sampled time series cannot
be easily generalized to unevenly spaced data. First we tried a simple but
robust statistic which measures the time reversibility : 
$\gamma\propto\sum[y_n-y_{n-1}/(t_n-t_{n-1})]^3$, where $y_n$ denotes the
count rate at time $t_n$ (e.g. Schmitz \& Schreiber 1999). For
a time series generated by a linear process, as in the case of surrogates,
we expect $\gamma\simeq 0$. Since the measure of time reversibility of a time
series generated by a nonlinear process can be either smaller or bigger than 
zero, we have performed a two--sided test. We found no statistically significant
differences between real data and surrogates for both the low and high count 
rate states, i.e., the analysis does not support the hypothesis of a non-linear 
time series.

However, it must be pointed out that asymmetry under time reversal is a 
sufficient but not a necessary condition for nonlinearity. In addition, it has 
been demonstrated (Schreiber \& Schmitz 1997) that, although the time 
reversibility is usually a good indicator of nonlinearity, in cases of very 
noisy data it can fail completely. According to Schreiber \& Schmitz (1997), 
who systematically evaluated the abilities of different observables to
detect nonlinearity, one of the best indicators is the nonlinear prediction 
error (e.g. Kantz \& Schreiber 1997), which relies on the time delay embedding 
of a scalar time series, described in section 5.3.

The results of a one--sided test (the nonlinear prediction error for a
nonlinear time series is expected to have lower values compared to the case
of a linear time series) for the high count rate state are shown in
Fig~\ref{figure:result2}, where the value of the nonlinear prediction error 
for the original data is plotted with a longer thick dashed line, the mean 
and the standard deviation of the statistic obtained from the surrogates are 
represented by a cross and an error bar, respectively. It is evident that the 
original data are singled out by the estimate of the nonlinear prediction 
error and, therefore, the null hypothesis of a linear stationary process 
has to be rejected at 97.5\% significance level.
Applying the same procedure to the interval of the light curve corresponding 
to the low state of \object{Ark~564}, we reach a similar conclusion: 
nonlinearity is detected, although at a slightly lower (95\%) significance level.

It must be noticed that also Edelson and collaborators (2002) searched for
nonlinearity in this source and claimed a weakly significant 
(less than 1.5 $\sigma$)  
detection of nonlinearity. However, as the authors themselves remark, the
method they used (the standard surrogate method) is not the best suited for
the ASCA data of \object{Ark~564}.

\begin{figure}
\psfig{figure=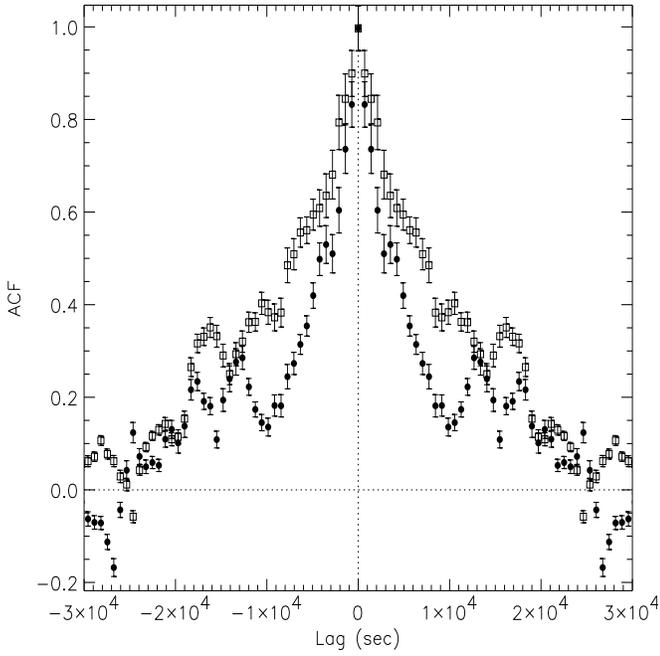,height=8.7cm,width=8.7cm,%
bbllx=50pt,bblly=122pt,bburx=592pt,bbury=600pt,angle=0,clip=}
\caption{Autocorrelation function of \object{Ark~564} during the high 
(filled circles) and low (open squares) count rate states. Note that the 
range of the horizontal axis is 
a factor five smaller than that in Fig.~\ref{figure:ACF-ab}. 
\label{figure:ACF-hl}}
\end{figure}

\begin{figure}
\psfig{figure=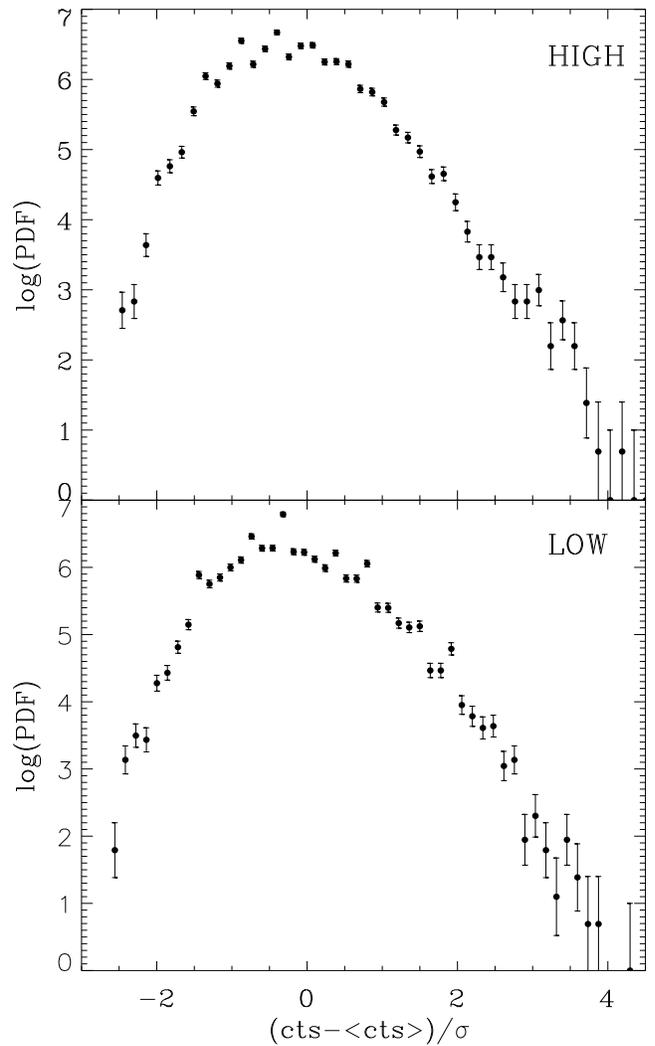,height=14cm,width=8.7cm,%
bbllx=80pt,bblly=40pt,bburx=397pt,bbury=567pt,angle=0,clip=}
\caption{Probability density function, normalized with respect to the average
count rate $<cts>$ and standard deviation $\sigma$ 
for time series in the high (top) and low (bottom panel) count rate states. 
 \label{figure:PDF}}
\end{figure}

\section{Time Series Analysis of \object{Ark~564}}
\subsection{Standard Analysis}
Detailed temporal and spectral variability analyses
of the 35 day ASCA observation of \object{Ark~564}, based on power density
spectra, cross--correlation functions are reported elsewhere
(e.g. Turner et al. 2001, Edelson et al. 2002, Papadakis et al. 2002) 
and are beyond the scope
of this work. Here, we investigate whether there are differences in the
variability behavior of the source during the low and high state. We use
standard techniques like the auto--correlation function and
excess variance analysis, linear techniques that are not 
frequently applied to AGN light curves
like the probability density function and, finally, fractal and nonlinear
analysis techniques.

An analysis of the discrete auto--correlation functions
(see Fig.~\ref{figure:ACF-hl}), 
for the two chosen intervals seems to indicate that the high state 
(filled circles) is characterized by a slightly shorter characteristic
 time scale
than the low count rate state (open squares), according to the narrower
 core of the former ACF.
 However, the most relevant quantity, the time lag at which the ACF decays to
zero, is the same ($\tau\sim2.5\times 10^4$ s) for both states.

To compare the degree of variability during the high and the low state
(Fig.~\ref{figure:lc2a})
we used the fractional variability amplitude $F_{\rm var}=\sqrt{(S^2-
\langle\sigma^2_{\rm err}\rangle)/\langle X \rangle^2}$, where $S^2$ is the total variance,
$\sigma^2_{\rm err}$ the mean error squared and $\langle X \rangle$ the mean count rate
(e.g. Edelson et al. 2002). We found that during the low state the variability
of the source was higher, $F_{\rm var}=(0.373\pm0.003)$, than 
during the high state $(0.300\pm0.002)$.
It is worth noticing that  from a recent temporal analysis based on 
$F_{\rm var}$ recently carried out on the ASCA light
curve of \object{Ark~564}, Turner et al. (2001) found no clear correlation 
between the fractional variability and the X-ray flux.

We have compared the probability density functions for the time series
of \object{Ark~564} normalized with respect to the average
count rate $<cts>$ and standard deviation $\sigma$
during the high and low state (see Fig~\ref{figure:PDF}).
The two distributions look quite similar and
differ from a Gaussian distribution (direct consequence of time 
correlations in the signal): both are characterized by a rather normal
left-hand tail (related to the small amplitude events in the time series) and a 
right-hand tail following a power law trend. It is worth noticing that a
qualitatively similar probability density function is displayed by two Galactic
accreting objects, the  micro-quasar \object{GRS~1915+105} and the black hole
candidate \object{Cygnus X-1} and not by a non--accreting object as
the Crab nebula (Greenhough et al. 2002).  

\subsection{Fractal Analysis}
\begin{figure}
\psfig{figure=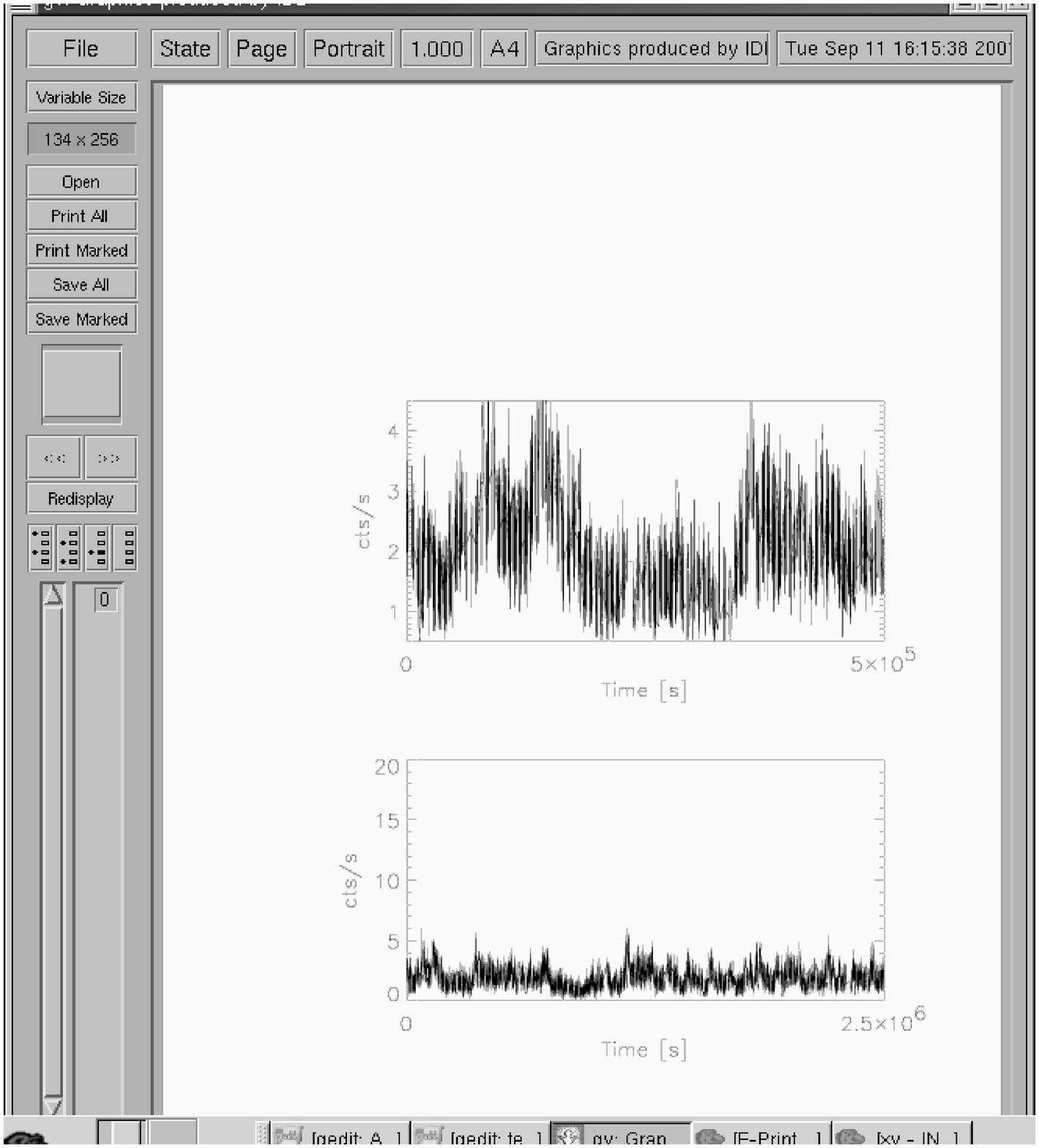,height=8.7cm,width=8.7cm,%
bbllx=192pt,bblly=-8pt,bburx=590pt,bbury=430pt,angle=0,clip=}
\caption{Interval of the light curve of \object{Ark~564} 
shown with high resolution (top panel) and low resolution (bottom). Both axes
are scaled by a factor of 5.
\label{figure:selfsim1}}
\end{figure}
Fractals are now widely used in the modeling and interpretation of
many different natural phenomena 
(see, e.g., Mandelbrot 1982, Pietronero \& Tosatti 1986), including
astrophysical phenomena (see, e.g., Heck \& Perdang 1991). 
In this section, we use methods tailored for fractal analysis
only as tools for
extracting information from the signal, without considering the dynamical
implications of the fractality. 
Our purpose is to go deeper into the comparison between the high 
and low count rate states, started in the previous section with standard 
techniques, by utilizing concepts characteristic of fractal analysis. In 
particular, after showing in a rather qualitative way that the light curve of 
\object{Ark~564} can be considered as a fractal object (characterized by
self-similarity and fractal dimension), we perform a quantitative comparison
of their fractal properties
based on the Hurst exponent (e.g. Kantz \& Schreiber 1997)

AGN light curves, sampled with sufficiently high resolution,
can be extremely complex and the scalar signal can be considered as a fractal,
in the sense that its graph, as a function of time, has a nontrivial 
dimension. This means that the length of the graph seen at a finite resolution
increases ``forever" when the resolution on the time axis is increased, since
more and more fluctuations are resolved.
A priory there should not be a minimum time scale of variability in an AGN, 
apart from causality reasons. For example, a process which is the result of
adding exponential shots will have a power density spectrum with a slope
$-2$ extending towards infinite positive frequencies.
Its dimension is, however, not an
integer but lies anywhere between one and two, depending on the resolution.
To visualize this concept let us consider a portion of the light curve 
of \object{Ark~564} scaled in time and space by the same factor (see 
Fig~\ref{figure:selfsim1}). In the limit of high resolution (top panel),
the graph tends to fill the plane and approaches dimension two,
whereas in the opposite limit, it degenerates towards a horizontal line and
thus to a dimension of one. Note that in this case the scaling factor is limited 
not by intrinsic physical reasons but by instrumental reasons (gaps due to 
earth occultation and low sensitivity). Moreover, approaching the 
resolution limit, the omnipresent noise component would hide the  small-scale
structures and the fluctuations would be washed out by measurement errors.

It must be noted that fractal properties, as the fractal length and
dimension of the X-ray light curve of an AGN (the Seyfert 2 galaxy 
\object{NGC~5506}) were already investigated by McHardy and Czerny (1987),
who derived some constraints on the emission mechanisms producing
X-rays. Here, our main aim is not to repeat a similar kind of analysis, but
to investigate the source's behavior at low and high
flux state, by comparing their fractal properties (i.e. their Hurst exponent,
see below).

\begin{figure}
\psfig{figure=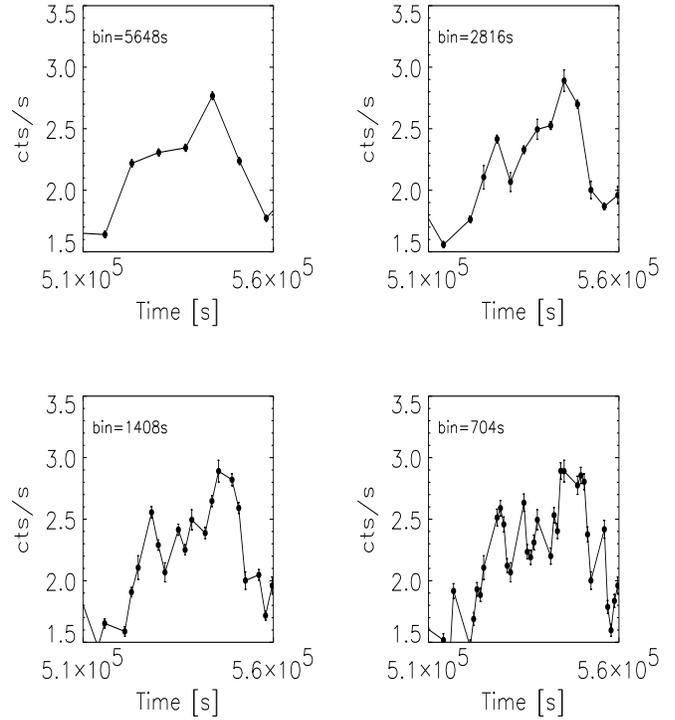,height=9.7cm,width=8.7cm,%
bbllx=60pt,bblly=95pt,bburx=545pt,bbury=520pt,angle=0,clip=}
\caption{Flare of \object{Ark~564} shown with increasing time resolution. 
The data points are connected with a line to show the averaged flare shapes.
\label{figure:selfsim2}}
\end{figure}

In addition to a nontrivial dimension, fractals are characterized
by self-similarity. A structure is said (strictly) self-similar if it can be
broken into arbitrarily small pieces, each of which is a small replica of the
entire structure. However, there are several variants of the mathematical 
definition of self--similarity. Dealing with erratic signals typical of AGN
X-ray light curves, we are mainly interested in the ``statistical
self-similarity" and self--affinity, where the small ``replica" may be 
somewhat distorted (for example skewed) with respect to the whole.

In the case of the light curve of \object{Ark~564}
this property can be evidenced by considering the same portion of the light
curve with different resolution factors (i.e. with different time binning).
In Fig~\ref{figure:selfsim2} we focus on one of the many flares  which occurred
during the ASCA observation of \object{Ark~564}. It is shown that a rather
smooth flare at low resolution (we started with a time binning equal to
the orbital period) looks more and more complex when seen with increasing 
time resolution: each single flare observed at higher resolution is composed
of several sub--flares. It can be seen  that from 1408 s  bins to
704 s bins  
the points start to cluster, indicating that the amount of variability on
time scales shorter than 1000s is smaller. This is also reflected in the
power density spectrum (Papadakis et al. 2002), which shows a break to a
steeper slope around $\nu=10^{-3}$ Hz.
\begin{figure}
\psfig{figure=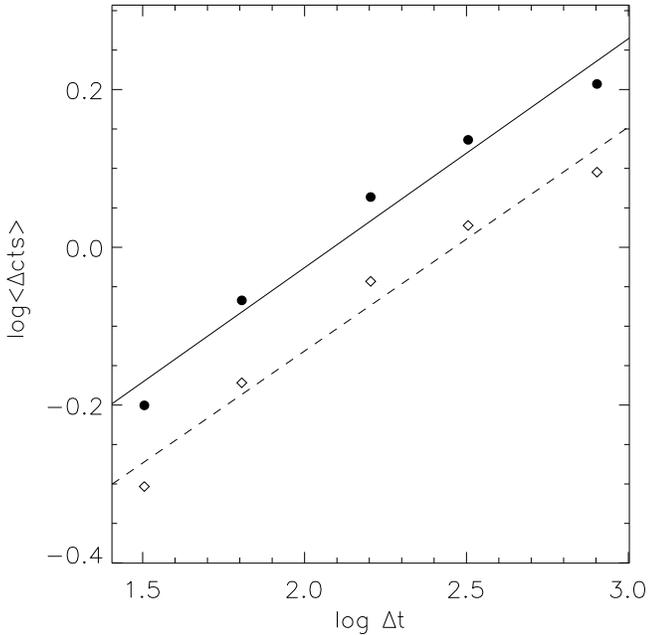,height=8.5cm,width=8.7cm,%
bbllx=47pt,bblly=180pt,bburx=467pt,bbury=598pt,angle=0,clip=}
\caption{Logarithmic plot of the growth of range $\langle \Delta cts \rangle$. 
The filled
circles refer to the high state while the diamonds to the low state. 
\label{figure:grow}}
\end{figure}
The frequent gaps caused by earth occultation
and the statistical fluctuations associated with measurement 
do not allow to use smaller time bins in the
scaling down process. It is worth noting that also
Tanihata et al. (\cite{tani}) noticed a similar behavior in a 
recent study on time variability of blazars: none of the observed large flares
show a smooth rise or decay but exhibit substructures, with smaller flares
having shorter time scales.

Statistical self-similarity, although only for a limited time interval 
(as indicated
by the breaks in the structure function and power density spectrum),
seems to occur independently of the state of 
the source. In order to investigate quantitatively 
the scaling behavior of the light curve on short time scales (which are not
accessible to visual inspection) during the high and low count rate states, 
we have  calculated,
for different time intervals $\Delta t$ (ranging from 32 to 1600 s),
the difference between the maximum and
the minimum count rates $\Delta cts$. Running this procedure over the portions of
the light curve corresponding to the high and the low states for each $\Delta t$,
an array of $\Delta cts$ is created from which the mean $\langle\Delta cts\rangle$ 
is found. For self--similar data $\langle\Delta cts
\rangle\propto
\Delta t^H$, where $H$ is called Hurst exponent (e.g. Kantz \& Schreiber 1997)
and ranges between 0 (for functions constant over time) and 1 
(for functions increasing or decreasing
linearly with time). Intermediate values of $H$ are generated by fractal
functions, random Gaussian noise ($H\simeq 0.2$) and Gaussian random walk
($H\simeq 0.5$; typical for Brownian motion and normal diffusion process),
whose increments possess a finite variance for every $\Delta t$ 
and are uncorrelated at successive time steps.

It is worth noticing that very different processes (e.g. hydrodynamical
turbulence, standard and anomalous diffusion) are able to produce
signals characterized by scaling laws and that they can be discriminated on
the basis of their Hurst exponent. For example, this method has been used
to quantify solar magnetic complexity (Adams et al. 1997), and the persistence
of solar activity (Lepreti et al. 2000). 
Fig~\ref{figure:grow} shows the growth of range of $\langle\Delta cts\rangle$ for 
\object{Ark~564} during the high (filled circles) and the low (open diamonds)
state. The slopes, obtained from a linear least square fit, are very similar
for the high ($0.29\pm 0.03$) and the low state ($0.28\pm 0.03$),
indicating that the processes at work during the high and low state are of the
same nature.

A similar analysis was performed by Greenhough and 
collaborators (2002) on very long RXTE light curves of three galactic objects: 
the \object{Crab} nebula, \object{Cygnus X-1} and the micro-quasar 
\object{GRS~1915+105}. Interestingly enough, only the latter, which is thought 
to be a scaled
down version of an AGN, had a Hurst exponent ($H\simeq 0.3$) consistent with
the values found for \object{Ark~564}.

\subsection{Nonlinear Analysis}
Linear methods intrinsically assume that the 
dynamics of the system is governed by the linear paradigm: small causes lead
to small effects. Since linear equations can only lead to exponentially
growing (or decaying) solutions or to periodic oscillations, an irregular
behavior of the system has to be attributed to some random external input.
However nonlinear, chaotic systems can produce very irregular data with
purely deterministic equations of motion. In other words, the apparently 
irregular and
unpredictable behavior typical of X-ray light curves of AGN is not necessarily
due to stochastic dynamics (i.e the action of a large number of excited 
degree of freedom). It might be also generated by chaotic dynamics of a limited
number of collective modes: dissipation, coupling between different degrees
of freedom and the action of an external field may lead to collective
behavior (e.g. Eckmann \& Ruelle 1985).
The presence of nonlinearity in the X--ray light curve of 
\object{Ark~564} leaves open the possibility that the light curve is produced
by a chaotic system. However, it must be noted that nonlinearity is a necessary
but not sufficient condition for a light curve to be produced by a chaotic
system (Vio et al. 1992). In any case our main aim is not to discriminate 
chaos from stochasticity, but to compare the source's behavior in the low 
and high state using nonlinear statistics. The scaling index method, the 
nonlinear statistics that we describe below, could be used even in the case of 
linear time series, since its ability to discern underlying structures in noisy 
data is not directly based on the nonlinear nature of the underlying process.

\begin{figure}
\psfig{figure=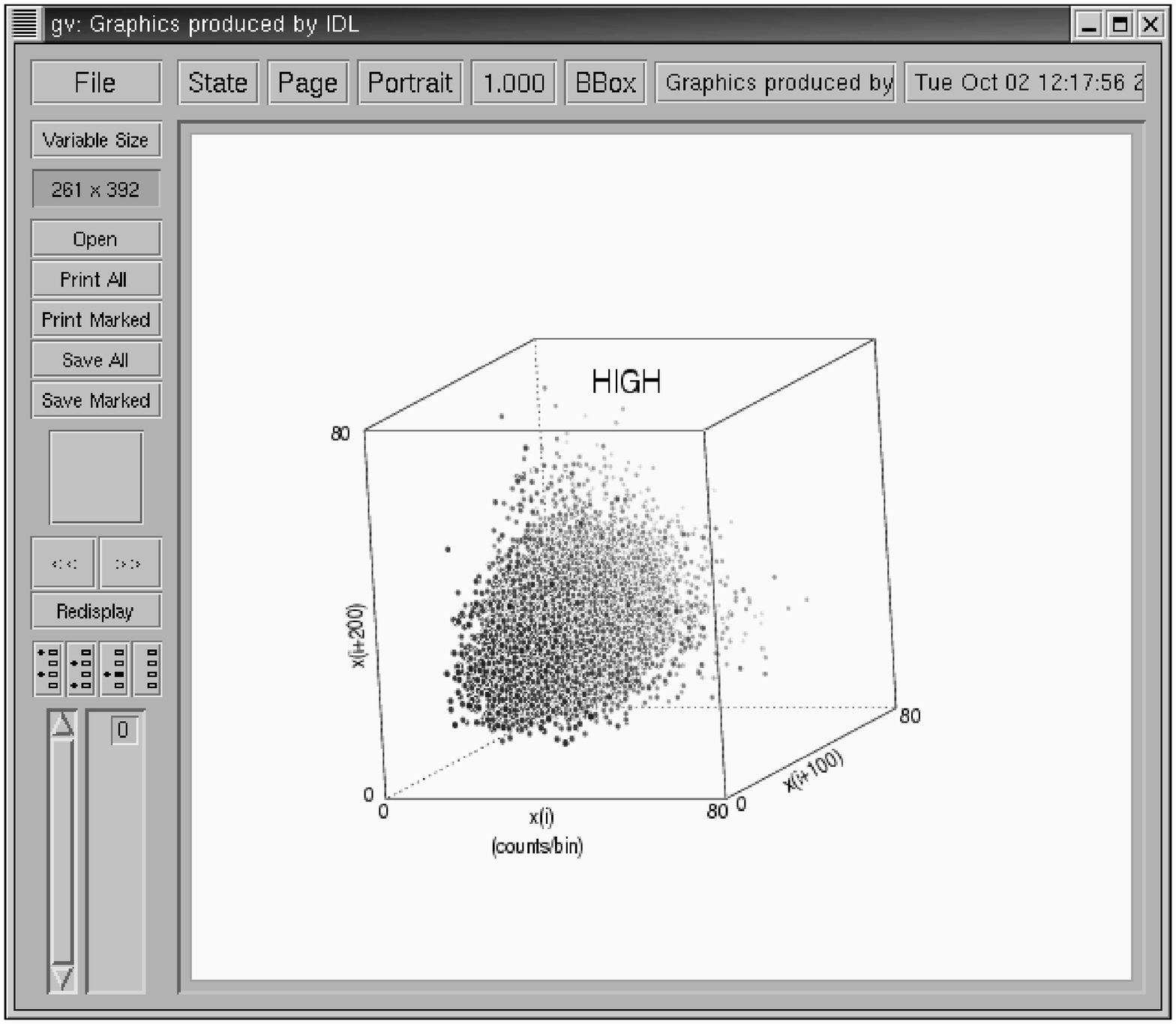,height=7cm,width=8cm,%
bbllx=165pt,bblly=203pt,bburx=482pt,bbury=499pt,angle=0,clip=}
\psfig{figure=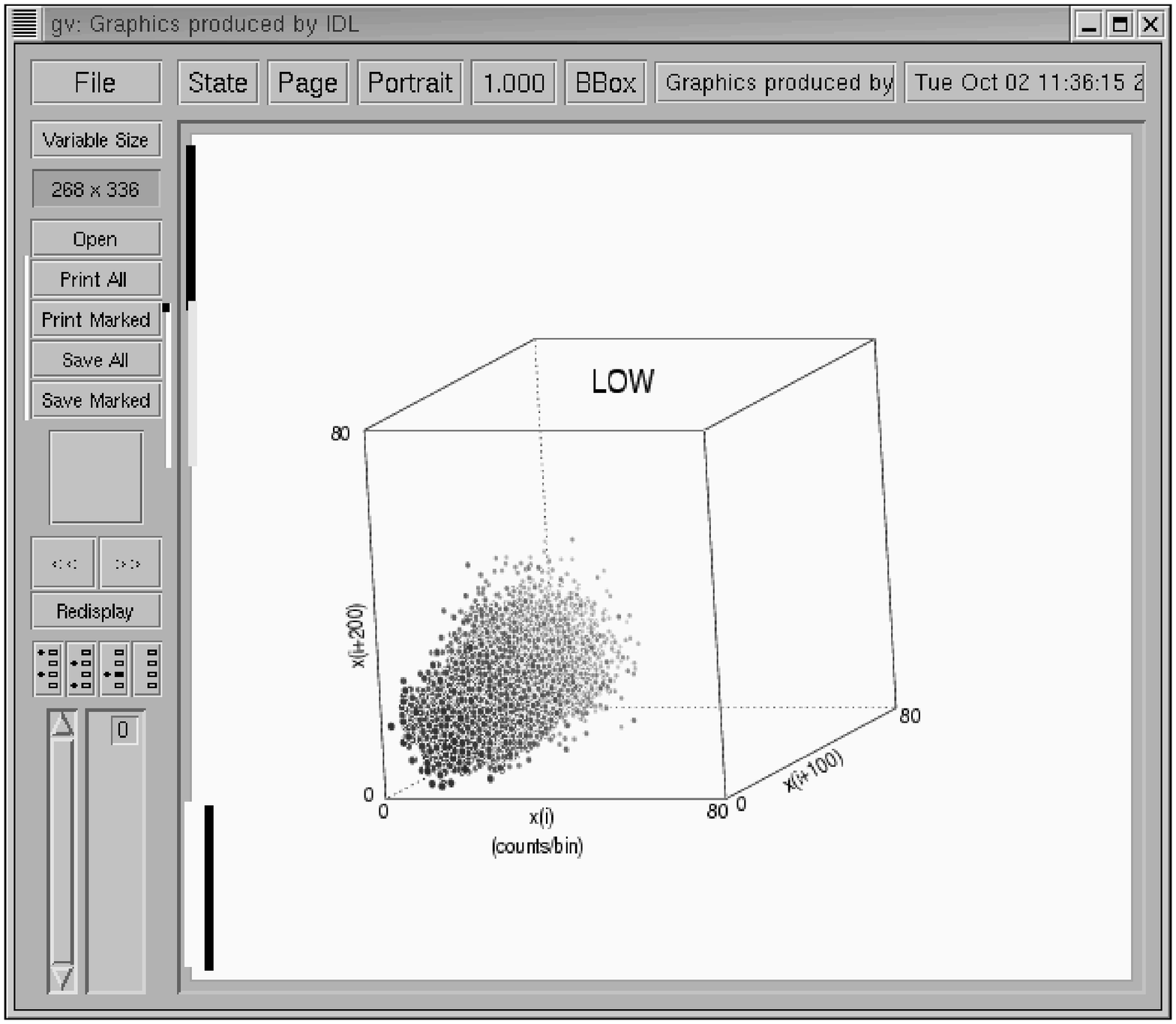,height=7cm,width=8cm,%
bbllx=165pt,bblly=203pt,bburx=482pt,bbury=499pt,angle=0,clip=}
\caption{3D--phase space portraits of the investigated portions of the light curve
of \object{Ark~564}, with 10000 data points each. The time delay used to
generate the artificial phase space is 100 times the temporal resolution of 
the time series. The units represent the number of photons contained in the
smallest time bin (i.e. in 16s).
\label{figure:attrac}}
\end{figure}

When methods of nonlinear dynamics are applied to time series analysis, the 
concept of phase
space reconstruction represents the basis for most of the analysis tools.
In fact, for studying chaotic deterministic systems it is 
important to establish a 
vector space (the phase space) such that specifying a point in this space
specifies a state of the system, and vice versa. However, in order to apply 
the concept of phase space to a time series, which is a scalar sequence of 
measurements, one has to convert it into a set of state vectors. This is
called phase space reconstruction and it is technically solved via the method
of time delays and embedding.
 Starting from the original time series $X[t_{\rm i}]$ and
introducing a time delay $\Delta t$, one can construct
$d-1$ additional data sets 
$X[t_{\rm i}+\Delta t]$,...,$X[t_{\rm i}+(d-1)\Delta t]$,
where $d$  is the dimension of the embedding space
(the so called``embedding dimension").
 The resulting $d$-dimensional artificial phase space represents
all topological properties of the system in the real phase space,
as long as the embedding dimension is twice the dimension
of the real phase space (Takens 1981).

For illustration purposes we show in Fig~\ref{figure:attrac} the 
three--dimensional pseudo phase space portraits of the light curve of 
\object{Ark~564} during the high (top panel) 
and the low (bottom) state. Roughly speaking, the location of each point in the 
3--D pseudo phase space (which is the simplest case, for a visual comparison 
between two data sets) is defined by three data points in the time series, which
are separated by 100 and 200 steps, respectively ($X_{i+100}$ means
an element displaced by 100 positions in the time series with respect $X_{i}$).
The time delay is almost arbitrary. However,
using too small time delays causes a clustering
of the state vectors around the diagonal (due to the strong time correlations
existing between successive elements of the vectors) and thus hamper the
discrimination between the two data set under examination. On the other hand, 
the choice of a large delay value would reduce the number of points in the phase
space, lowering the statistical significance of the analysis.
Note that discontinuities in the time series due to earth occultation do not
represent a problem for this kind of analysis, as far as the  
number of data points is sufficiently large for the  reconstruction.

\begin{figure}
\psfig{figure=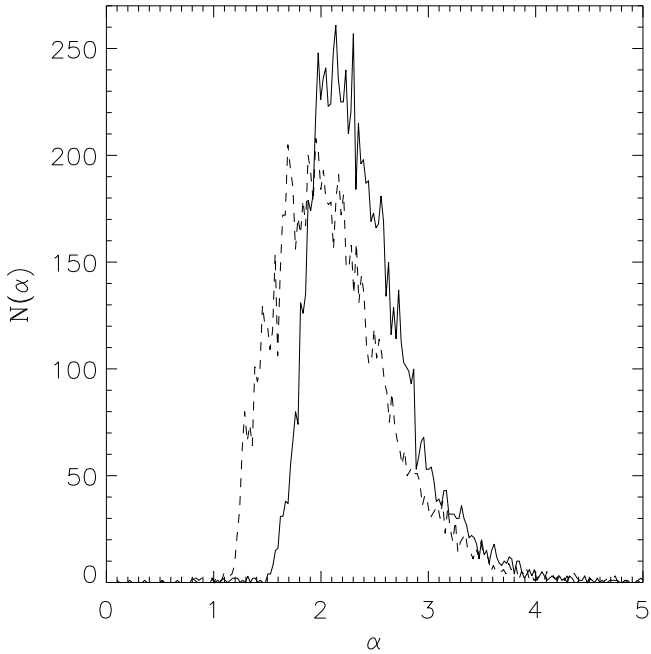,height=8.7cm,width=8.7cm,%
bbllx=30pt,bblly=15pt,bburx=435pt,bbury=429pt,angle=0,clip=}
\caption{Spectrum of the scaling index distribution of \object{Ark~564}
during the high (solid line) and low (dashed line) states.
\label{figure:f-alpha}}
\end{figure}
 
In order to quantify the difference between the phase space 
portraits in the high and low states, a suitable quantity is the correlation 
integral $C(r)$, which basically counts the number of pairs of points with 
distances smaller than $r$. This algorithm was introduced by Grassberger 
\& Procaccia (1983) and was first used in astronomy to analyze light curves of 
variable stars (Auvergne \& Baglin 1986) and binary systems (e.g. Voges et al.
1987), in order to put constraints on the origin of the irregular variability.
An exhaustive description of the correlation integral method and its application
to X--ray light curves of AGN is given by Lehto and collaborators (1993). 
An important property is that $C(r)$ at small r behaves as a power law and the exponent 
is closely related to the correlation dimension $D^{(2)}$ (Grassberger \& 
Procaccia 1983), which for deterministic dynamical systems provides an estimate 
of the number of the degree of freedom excited, and more 
generally gives an indication of the ``complexity" of the system.

The scaling index method (e.g. Atmanspacher et al. \cite{atma}), which is based 
on the local estimate of the correlation integral for each point in the phase 
space,
has been employed successfully in very different fields of the science for its
ability to discern underlying structure in noisy data. A detailed description
of this method and its application to astrophysics has been given by
Williams et al. (\cite{willi}). In practice, the scaling index method 
characterizes quantitatively the data point distribution 
by estimating the ``crowding" of the data around each data point. For each of the
N points, the cumulative number function is calculated
\begin{equation} 
N_i(r)=\#\{j\vert d_{ij}\leq r\},
\end{equation}
where $\#\{j\}$ indicates the number of points $j$, whose distance 
$d_{ij}$ from a point $i$ is smaller than $r$. The function $N_i(r)$ for each 
$i$ in a given range of radii (which are related to the typical distances 
between the data points, that, in turn, depend on the choice of the embedding 
space dimension) is approximated with a power law
\begin{equation} 
N_i(r)\sim r^{\alpha_i}\,\,\,(r_1<r<r_2),
\end{equation}
where $\alpha_i$ are the scaling indices. Explicitly, the $\alpha_i$ are given by
\begin{equation} 
{\alpha_i}=\frac{\log N_i(r_1)-\log N_i(r_2)}{(\log r_1-\log r_2)},
\end{equation}
Note that for a purely random process the average scaling index tends to the 
value of the embedding dimension, whereas for regular and for deterministic 
(chaotic) processes the value of $\langle\alpha\rangle$ is always smaller 
than the dimension of the embedding space.

It has been pointed out by Vio and collaborators (1992)
that a naive application of the correlation integral
method may lead to conceptual errors (as the detection of low--dimensional
dynamics from signals produced by stochastic processes) and that this method
must not be used in isolation, but as one of a collection of tests to
discriminate chaos from stochasticity. Here we stress once more that
the nonlinear statistics based on the scaling index is used 
only as a statistical test to discriminate between two states
of the system, not to determine any kind of dimension, which would be meaningless
for non-deterministic systems. The only requirement for the nonlinear 
discrimination of time series data, as for any statistical test, is that it
must yield significant results. 

We have calculated the scaling index for all 
the points of the two selected parts in the light curves using an embedding 
space of dimension 4 and suitable radii. There are no strong prescriptions 
for the
choice of the embedding dimension. The discriminating power of the statistic 
based on the scaling index is enhanced using high embedding dimensions (this 
can be easily understood in the following way: if the data are embedded in a 
low-dimension phase-space many of them will fall on the same position, losing 
in this way part of the information). On the other hand, the choice of too 
high embedding dimensions would reduce the number of points in the
pseudo phase-space, lowering the statistical significance of the test. 
We performed the scaling index analysis by increasing systematically the 
embedding dimension from 2 to 10, and found that a dimension of 4 is the 
best choice for the two investigated portions of the light curve
with 10,000 data points each.  The choice of the radii is also rather
arbitrary, provided that Eq. (3) holds. 

The resulting histograms are plotted in 
Fig~\ref{figure:f-alpha}.
There is a remarkable difference between the two distributions,
both in shape and location of the centroid. 
Such a difference cannot be ascribed to the 
fact that the data point distribution in the low state occupies a 
lower region in the pseudo phase-space with respect to the 
high space distribution. In fact the scaling
index analysis is sensitive only to the relative positions of the data 
points within a distribution, not to their absolute positions in the phase 
space. For this kind of analysis the most important quantity is the location 
of the maximum of the histogram, which is related to the correlation dimension. 
Bearing in mind that the correlation dimension of a
purely random data distribution tends to the value of the embedding dimension, 
the location of the maximum in the high state at a larger value indicates 
that the high state is ``more random" or more ``complex" (using a term 
borrowed from nonlinear dynamics) than that of the low count rate state. 
In other words, the high state is characterized by the action 
of a larger number of excited degrees of freedom. 

\section{Conclusions}
We have carried out a thorough analysis of the timing properties of the
NLS1 \object{Ark~564}, using the data from a recent long observation 
performed by the ASCA satellite. 

The first important result is that we searched for signs of non--stationarity,
without finding any strong evidence for it, although the differences in the 
structure function and autocorrelation functions between the first and the 
second part of the light curve give some indications in favor of a 
non--stationarity of the total light curve. To asses
the stationarity of a time series is also important for a different aspect:
the non-stationarity of the signal can strongly affect the identification
of genuine nonlinearity possibly present in the data, and lead to wrong
interpretations of results obtained from nonlinear statistical analyses.

As sensitive light curves long enough to cover several
transitions between high and low states are presently not available,
we concentrated on two sub--intervals
of the \object{Ark~564} light curve, which are locally stationary, contain
a sufficiently large number of data points (10000, using a bin size of 
16 s) for a meaningful statistical analysis, and which have significantly 
different
mean count rates in order to characterize the intrinsic temporal differences
from the low and the high state of the source.

We then carefully addressed the issue of the possible presence of nonlinearity,
(often invoked as cause of giant flares in AGN light curves), which is crucial
for breaking the degeneracy of models able to reproduce time-averaged spectra
and some general timing properties. The result, obtained by utilizing a 
generalized surrogate method (well suited for unevenly sampled data)
and the nonlinear prediction error, indicates the presence of nonlinearity
both in the high and, at a slightly lower significance level, in the low 
count rate states. As a consequence, intrinsically linear models, 
where the variability is caused by many independent active regions, as 
magnetic flares or the traditional ``shot noise"
should be ruled out for \object{Ark~564}. Generalizations of these models,
where the spatial and temporal distributions of flares are not random (e.g.
Merloni \& Fabian 2001) might still be a viable solution. The presence of 
nonlinearity favors nonlinear models as the self-organized criticality model 
or the emission from the base of a putative X-ray jet, as recently found in 
several radio-loud AGN. This last hypothesis seems to be  
supported also by two independent arguments: 
1) the need of relativistic beaming effects to explain the
extreme values for the efficiency in the conversion of gravitational potential 
energy into X-ray emission in several NLS1 galaxies (e.g. \object{PKS~0558-504},
Remillard et al.(1991); \object{PHL~1092}, Brandt et al. (1999); 
\object{RX J1702.5+3247}, Gliozzi et al.(2001)). 2) The recent discovery that, 
in a sample of 62 NLS1 detected in the FIRST VLA radio survey 
(Becker \cite{beck}), $\sim 40\%$ of the objects are radio--loud ($R>10$) and 
the remainder fall in the radio-intermediate range ($1<R<10$) 
(Whalen et al. 2002). 

We then tried to characterize the timing behavior during the low and high
count rate states of \object{Ark~564}. First, we used
standard techniques (like the auto--correlation function and
excess variance analysis) and linear techniques that are not frequently 
applied to AGN light curves (like the probability density function), without 
finding any significant difference between the two states. After showing that, 
for a limited time interval, the complex light curve
of \object{Ark~564} might be viewed as fractal objects with a non--trivial
fractal dimension and statistical self--similarity, we tried to discriminate 
the two states with the help of the Hurst exponent. 
Also with this analysis no intrinsic differences were found, leading to the 
conclusion that the physical processes during the 
high and low state are of the same nature.

Finally, we have introduced the scaling index spectrum, based on phase space
reconstruction, as a tool to discriminate time series. Based on this
nonlinear statistic, we showed that the low and high states of \object{Ark~564},
apparently quite similar with respect to their linear properties, are 
intrinsically different, with the high state characterized by a higher degree 
of complexity, meaning that the number of
degrees of freedom during the high state is larger that during the
low state. 
However, in order to obtain firmer and more general conclusions we have to 
await further observations, in particular from the new X--ray missions, with 
sensitive, high time resolution instruments and the possibility of long 
continuous observations.
                     
\begin{acknowledgements} 
We thank the anonymous referee for the useful comments and suggestions that
improved the paper. 
MG acknowledges financial support by NASA LTSA grant NAG5-10708
Part of this work was done in the TMR network ``Accretion onto black holes,
compact stars and protostars", funded by 
the European Commission under contract number ERBFMRX-CT98-0195 
\end{acknowledgements}

\end{document}